\documentclass[12pt]{report}
\input{psfig}
\begin{document}

\begin{titlepage}
$\;$\vspace{1cm}
\vspace{1cm}
\begin{center}{\Huge  {Two Potential Mechanisms of Spatial Attention in Early Visual Areas}} \end{center}
\vspace{1cm}
\begin{center} {\large { \bf by Erez PERSI}} \end{center}
\vspace{1.5cm}
\begin{center} {\large { \bf Laboratoire de Neurophysique et Neurophysiologie, CNRS UMR8119, Universit\'e Paris Descartes, 75270, Paris Cedex 06, France}} \end{center}
\begin{flushleft} 
\Large{ Correspondence should be addressed to: \\
erez.persi@univ-paris5.fr
}
\end{flushleft}
\end{titlepage}

\baselineskip 18pt
\bibliographystyle{unsrt}
\pagestyle{plain}

\begin{abstract}

   We investigate theoretically the effect of spatial attention on the 
   contrast-response function (CRF) and orientation-tuning curves in early visual areas.
   We look at a model of a hypercolumn developed recently (Persi et al., 2008), 
   that accounts for both the contrast response and tuning properties in the primary visual cortex, 
   and extend it to two visual areas.
   The effect of spatial attention is studied in a model of two inter-connected visual areas,
   under two hypothesis that do not  necessarily contradict.
   The first hypothesis is that attention alters inter-areal feedback synaptic strength,
   as has been proposed by many previous studies.
   A second new hypothesis is that attention effectively alters single neuron input-output properties.
   We show that with both mechanisms it is possible to achieve attentional effects similarly
   to those observed in experiments, namely contrast-gain and response-gain effects, while keeping the
   orientation-tuning curves width approximately contrast-invariant and attention-invariant.
   Nevertheless, some differences occur and are discussed.
   We propose a simple test on existing data based on the second hypothesis.

\end{abstract}

\newpage
 
   \section*{Introduction}

   The effects of spatial attention on the responses of neurons in the brain are numerous.
   The nature of the observed attentional effects depend on many factors,
   such as the number of stimuli, their optimality, their spatial distribution, the task and the cue. 
   This as a consequence often leads to a considerable confusion and difficulty in our understanding of the 
   underlying mechanisms by which attention acts.
   To elaborate on the possible mechanisms,
   we focus here on the effects of attention on the CRF and orientation tuning curves 
   that have been observed in a simple 'In-Out' configuration, 
   where one stimulus is presented inside the receptive field (RF) of a neuron, 
   and a second stimulus appears outside the neuron's RF, 
   such that sensory interaction between the stimuli is weak.
   'Attention' is defined as the difference between the response of the neuron when 
   the subject attended to the stimulus in the RF (attended-mode),
   and the response of the neuron when the subject attended the stimulus outside the RF.\\

   \emph{Multiplicative effects}:\\
   Two types of multiplicative effects have been observed in the CRFs of neurons in V1-V4 
   (McAdams \& Maunsell 1999; Reynolds et al, 2000; Williford \& Maunsell, 2006): 
   1) An effective change in stimulus contrast, named a contrast-gain effect. 
   Here, attention leads to a leftward shift of the CRF, thus the effect at low and high contrasts is weak.
   2) An effective change in the response, named a response-gain effect.
   Here, attention leads to an upward shift of the CRFs, thus the effect is weak only at low contrasts.

   In McAdams \& Maunsell (1999) the orientation-tuning curves of neurons 
   in V1 and V4 were measured with stimuli of changing contrast. 
   Thus, the stimulus contrast was not well-defined.
   In V4, 85\% of the cells were affected by orientation, 55\% were affected by attention, and 47\% by both.
   Out of the 47\%, 86\% showed an average increase of 26\% in the firing rates in the attended mode, 
   and across all orientations in a multiplicative fashion.
   In V1, 99\% of the neurons were affected by orientation but only 31\% by attention.
   Attention, had a less significant effect of an average 8\% increase in the firing rates.
   In all areas, orientation-tuning width was approximately attention-invariant, 
   indicating that orientation-tuning and attention do not interact.
	
   To test if this observation is consistent with either contrast or response multiplicatives,
   Reynolds et al. (2000) used drifting gratings of several contrasts that spanned the CRF of neuron's in V4.
   In the attended mode the average CRF of neurons was shifted to the left, saturation level did not change,
   an indication of a contrast-gain mechanism, but larger effects were measured at low contrasts.
   Further evidence for a contrast-gain effect comes from the study of Fries et al. (2001) who used high-contrast 
   stimuli and observed no change in activity, but only a change in the power spectrum of the 
   spike-triggered averages (STAs) of the local field potential. 
   Also in Luck et al. (1997) and Reynolds et al. (1999) 
   attention had no consistent effect besides a change in the baseline activity for high-contrast stimuli,
   for the 'In-Out' configuration.
   However, in a later study (Williford \& Maunsell, 2006), changes in the CRF of individual neurons showed a large diversity.
   CRFs could be well fitted to either a contrast-gain leftward shift, or to a response-gain upward shift.
   The differences in the results between these studies remain unclear,
   despite a recent attempt to resolve this discrepancy using fMRI (Boynton \& Buracas, 2007).\\

   \emph{Signal initiation}:\\
   In theory, attention can be either initiated in low cortical areas that transmit a bottom-up feedforward signal,
   or can be initiated in higher cortical areas and transmit a top-down feedback signal.
   Recent studies from both single-cell physiology and neuropathology provide strong evidence for an initial 
   activation of fronto-parietal regions that serve a top-down signal (for review, see Pessoa et al, 2003).
   Perhaps the most striking evidence comes from fMRI studies (Kastner et al, 1998; Hopfinger et al., 2000),
   showing that these areas, as well as some temporal areas, 
   are activated already in the presence of a cue to a spatial location, before any sensory stimulus appears.
   The increased activity in these areas was sustained during the stimuli presentation,
   reflecting operation of the task and not of sensory processing.
   The 'target' activated bilaterally the supplementary motor, ventrolateral and prefrontal areas.
   In electrophysiology studies the reaction to the cue, during the expectation to the stimulus, 
   increased the baseline activity by 40\% (Luck et al, 1997; Reynolds et al, 1999; Reynolds \& Pasternak, 2000). 
   Since, this did not modulate spike activity, it indicates a signal which is affected only by attention. 
   The functional role of such a signal is unknown.
   For the processing of the sensory input the effect of attention 
   in Kastner et al. (1998) was larger for higher visual areas, in agreement with electrophysiology.

   \section*{A model of multiplicative effects of spatial attention}

   We focus on the multiplicative effects on the CRFs and orientation tuning curves observed in the 'In-Out' configuration, 
   because it represents the simplest manipulation that avoids the influence of sensory interaction among stimuli
   (Luck et al., 1997; Reynolds et al., 1999).
   We first describe the essential properties of a V1 hypercolumn model, 
   described by a ring architecture (Ben-Yishai et al., 1995), composed of both excitatory ($E$) and inhibitory ($I$) neurons,
   whose transfer-functions are nonlinear.
   The model has been described in detail in Persi et al. (2008),
   where we demonstrated that the hypercolumn behaves similarly with a rate neuron model and a conductance-based neuron model.
   Therefore, here, we briefly describe the contrast and tuning properties of rate model neurons
   in a hypercolumn model which will serve as a reference model for a single visual area.
   Then, we extend the model to two visual areas (for example V1-V2) and examine two potential mechanisms,
   that attention either leads to changes in inter-areal feedback strength or leads to changes 
   in the effective single-neuron transfer-function properties.
   Under the assumption of a top-down attentional signal we assume that these changes originate in the higher area.
   The goal is to find under which conditions multiplicative contrast-gain and response-gain effects are observed,
   such that the width of the orientation-tuning curves remain contrast and attention-invariant. 
	
        \subsection*{Model of a single visual area}
\begin{figure}[btp]
	  \centerline{\psfig{figure= 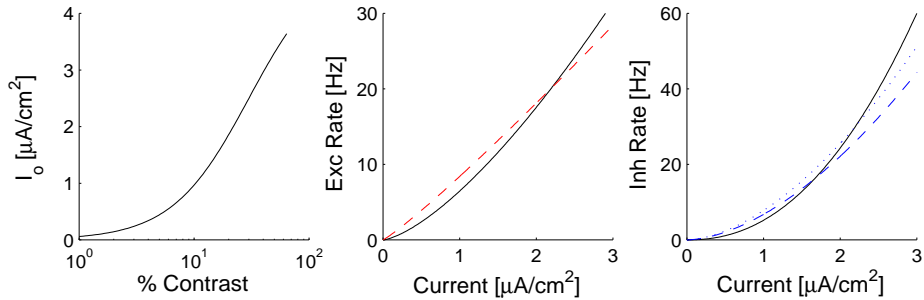,width=350pt}}
	  \caption[]{ Left) The input-contrast H-ration relationship, for contrast percentage interval [1,64], 
	    with \(I_{max} = 5\mu A/cm^{2}\), \(n = 1.3 \) and \(C_{50} = 30\% \). 
	    Middle) The excitatory power-law input-output relationship for the reference set 
	    \( \beta_{E} = 6.5,\  \alpha_{E} = 1.45 \) (solid);  
	    \( \beta_{E} \to 1.3\beta_{E}, \ \alpha_{E} \to \alpha_{E}/1.3 \) (dashed).
	    Right) The inhibitory transfer-function for the reference set 
	    \( \beta_{I} = 5.2,\  \alpha_{I} = 2.2 \) (solid);
	    \( \beta_{I} \to 1.3\beta_{I}, \ \alpha_{I} \to \alpha_{I}/1.3 \) (dashed).
	    \( \beta_{I} \to 1.5\beta_{I}, \ \alpha_{I} \to \alpha_{I}/1.3 \) (dotted).
	    The CRFs of excitatory neurons and inhibitory neurons typically operate in the range 
	    of 0-20Hz and 0-40HZ, respectively (below).
	    For our choice of parameters, a larger change in $\beta$ than in $\alpha$ is needed 
	    for the inhibitory to operate in their range, whereas for the excitatory neurons 
	    same changes in $\alpha$, $\beta$ are enough.
	  }
	  \label{RIC}
\end{figure}

	The preferred orientation $\theta$ of both neuron types is uniformly distributed over the interval \([-\pi/2, \pi/2]\).
	The neurons' nonlinear input-output transfer-function is assumed to be well described by a power-law function
	(Anderson et al., 2000; Priebe et al., 2004), a consequence of noise in the inputs 
	(Hansel and van Vreeswijk, 2002; Miller and Troyer, 2002).
	Therefore, the steady state firing rate $R_{a}$ of a neuron of type \( a = E, \ I \) 
	as a function of the input current $I_{a}$ is given by: 
	\begin{equation}
	  R_{a}(I) = \beta[I_{a} - I_{th}]_{+}^{\alpha_{a}}
	  \label{pl}
	\end{equation}
	The function's three constants are the gain $\beta$, 
	the power-law exponent $\alpha$ and the effective current threshold $I_{th}$.
	We assume that inhibitory neurons have larger firing rates than excitatory neurons, 
	such that \( \alpha_{I} > \alpha_{E} \) (figure \ref{RIC}, middle and right), 
	and that noise effectively erases the effect of the spiking threshold such that 
	for both neuron types \(I_{th} = 0\).

	As a function of orientation, a neuron receives a Gaussian input from the LGN.
	Without any loss of generality one can assume that the input peaks at \(\theta = 0\), thus:
	\begin{equation}
	  I_{a,LGN}(\theta,C) = I_{0}(C)\exp(-\theta^{2}/2\sigma_{a,LGN}^{2})
	\end{equation} 
	where $C$ is the contrast percentage, and $\sigma_{a,LGN}$ is the LGN input width.
	LGN cells' firing rates are well fitted to an H-ratio function 
	(Albrecht and Hamilton, 1982; Sclar et al., 1990),
	thus we describe the current amplitude by:
	\begin{equation}
	  I_{o}(C) = I_{max}\frac{C^{n}}{C_{50}^n + C^{n}}
	\end{equation} 
	where $I_{max}$ is the maximum current amplitude and $C_{50}$ is the contrast at which the \(I_{0} = I_{max}/2\).
	The function is shown in figure \ref{RIC} (left).

\begin{figure}[btp]
	  \centerline{\psfig{figure= 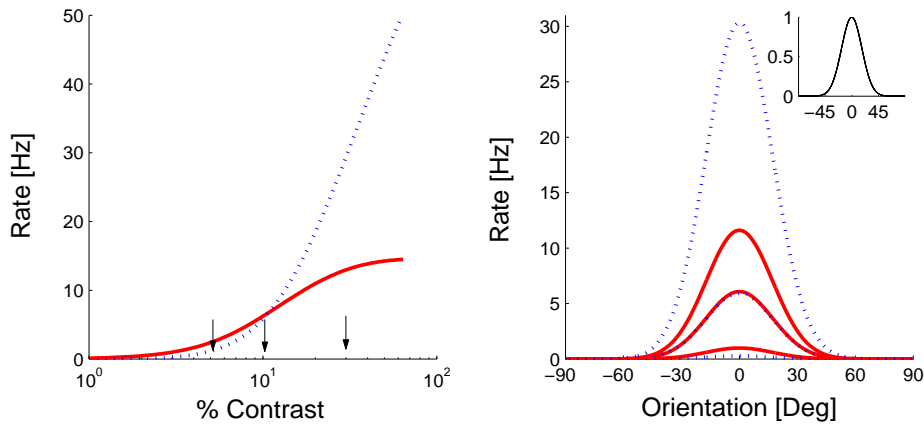,width=350pt}}
	  \caption[]{ Left) The CRFs of excitatory neurons (solid) and inhibitory neurons (dotted).
	    Synaptic strength parameters are \( J_{EE} = 0.06, \ J_{EI} = -0.0625, \ J_{IE} = 0.06, \ J_{II} = -0.0435 \).
	    LGN input and transfer-function parameters are those of the reference set (figure \ref{RIC}, solid).
	    The arrows point to contrast value of 3\%, 10\% and 30\%.
	    Right) The orientation-tuning curves for these three levels of contrast. 
	    The inset shows all the orientation tuning curves when normalized to the peak response.
	    Because of the inherent orientation-tuning width invariance, 
	    and because we chose an equal tuning width for the excitatory and inhibitory neurons, all curves coincide.
	  }
	  \label{CRFsOTs}
\end{figure}

	The recurrent connections from post-synaptic neurons $b$ of preferred orientation $\theta'$
	to pre-synaptic neurons $a$ of preferred orientation $\theta$ are also Gaussian,
	\(J_{ab}(\theta,\theta') = J_{ab}/(\sigma_{ab}\sqrt{2\pi})\exp[-(\theta - \theta')/2\sigma_{ab}^2]\).
	The width $\sigma_{ab}$ are such that all three input sources, 
	LGN, excitatory and inhibitory to a given neuron have the same width, thus their sum is also a Gaussian of the same width.
	Under these conditions: \(\sigma_{a,LGN}^2 = \sigma_{aE}^2 + \sigma_{E}^{2} = \sigma_{aI}^{2} + \sigma_{I}^{2} \),
	the firing rate self-consistent solution is a Gaussian with a constant width $\sigma_{a}^{2}$,
	which is narrower than the input width by the factor $\sqrt{\alpha}$ for \( \alpha > 1\).
	All inputs and the rate orientation-tuning width are contrast-invariant (Sclar et al., 1982; Skottun et al., 1987).
	The case for which excitatory and inhibitory neurons have the same orientation-tuning width
	is demonstrated in figure \ref{CRFsOTs} (right), for the LGN input shown in figure \ref{RIC} (left).

	In figure \ref{CRFsOTs} (left) we demonstrate that the sigmoid shape of the contrast-response functions (CRFs) is captured,
	with acceleration at low contrast and saturation for high contrast 
	(Albrecht et al., 1982; Sclar et al., 1990; Contreras \& Palmer, 2003).
	This behavior is robust for our choice  of \( \alpha_{I} > \alpha_{E} \), an inequality that we found also in experimental data.
	Acceleration of both neuron types originates in the nonlinearity of neurons' transfer-functions.
	The shape of the CRFs at high contrast is different for the two neuron types.
	For excitatory neurons, contrast saturation is strongly affected by the recurrent inhibition at high contrasts,
	while inhibitory neurons follow the saturating properties of their input.
	The inhibitory recurrent input therefore leads to an early saturation of
	excitatory neurons with steeper CRF than their LGN inputs in agreement with 
	Sclar et al. (1990) and Finn et al. (2007).
	In Persi et al. (2008) we show that the CRF shape mainly depends on the ratio of the inhibitory
	input, \( J_{EI}/J_{II} \), where $J_{EE}$ scales the CRF in a response-gain fashion. 
	Because of the orientation-tuning width invariance, the CRFs of neurons of different
	preferred orientation are similar up to a constant factor of \(\exp(-\theta^{2}/2\sigma_{a}^{2})\), 
	in agreement with experimental observations  (Albrecht et al., 1982; Skottun et al., 1987).

	\subsection*{Behavior of two visual areas}
	We assume that two identical rings of the type described above represent two visual areas, for example, V1-V2.
	Therefore, each visual area receives not only recurrent inputs but also inter-areal feedback inputs.
	In general, the total current to the soma of a neuron of type $a$ in layer $k$ is:
	\begin{equation}
	  I_{a}^{k}(\theta) = \int_{-\pi/2}^{\pi/2} \sum_{b=E,I}\sum_{k,l=1}^{2} J_{ab}^{kl}(\theta - \theta^{'})R_{b}^{l}(\theta^{'})d\theta^{'} +
                      I_{a,LGN}^{1}(\theta).
		      \label{Iatt}
	\end{equation}
	However, we will assume that inter-areal long-range synaptic connections can be only excitatory, 
	\( J_{aI}^{kl} = 0 \), for \( k\neq l \).
	We further assume that the excitatory inter-areal synaptic profiles $J_{aE}^{kl}(\theta,\theta')$, \( k \neq l\), 
	obey the conditions for contrast invariance input tuning width,
	similarly to the recurrent connection profiles $J_{ab}^{kl}(\theta,\theta')$, \( k = l\).
	Therefore, any change in the CRFs due to changes in the inter-areal feedback strength is consistent 
	with contrast-invariant orientation-tuning width.
	
\begin{figure}[btp]
	  \centerline{\psfig{figure= 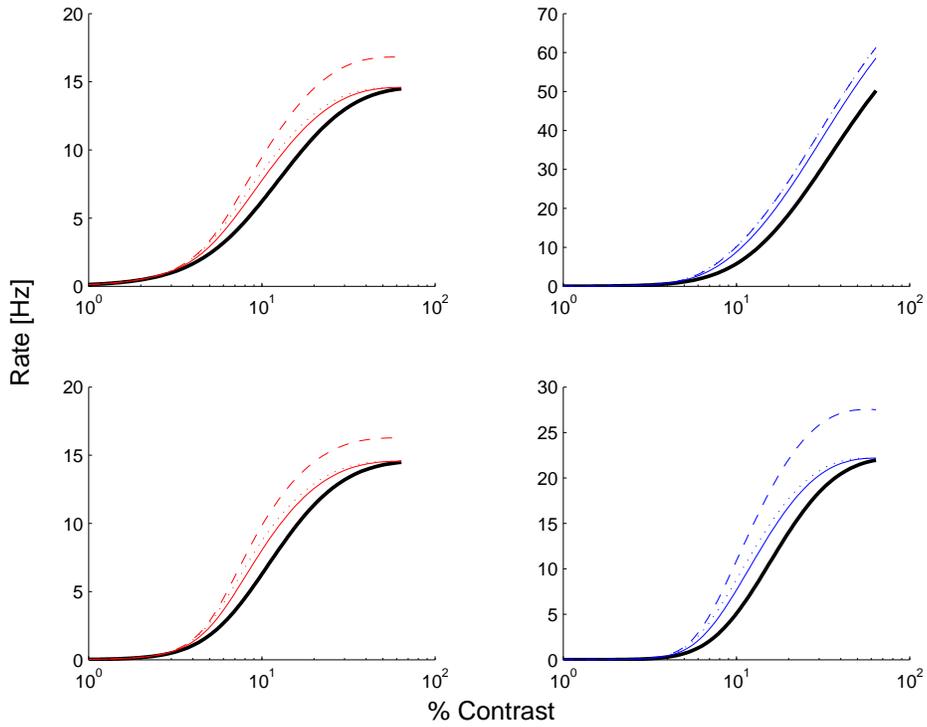,width=350pt}}
	  \caption[]{ The effect of changes in inter-areal feedback strength. 
	    In the upper panel the CRFs of excitatory (right) and inhibitory (left) in V1. In the lower panel the CRFs of V2 neurons.
	    Solid think curves represent the activity in the system with only feedforward connection: LGN-V1-V2.
	    The LGN input is give in figure \ref{RIC}, the feedforward strength are \( J_{EE}^{21} = J_{IE}^{21} = 0.15 \).
	    Solid thin curves are the CRFs when feedback connections are added with \( J_{EE}^{12} = J_{IE}^{12} = 0.03 \).
	    Dotted curves are CRFs when the feedback strength are equally enhanced through the attentional signal 
	    with \( J_{EE}^{12} = J_{IE}^{12} = 0.04 \). Dashed curves represents a larger increase in $J_{EE}^{12}$ with
	    \( J_{EE}^{12} = 0.04, \ J_{IE}^{12} = 0.03 \).
	  }
	  \label{FBeffect}
\end{figure}

	For the inter-areal feedforward stream, we assume that V1 receives LGN input,
	and V2 receives V1's output, where
	$J_{EE}^{21}$ and $J_{IE}^{21}$ are chosen such that the activities in both areas 
	fairly represent electrophysiological measurements.
	CRFs of both neuron types for this LGN-V1-V2 stream, without inter-areal feedback from V2 to V1, 
	are shown in figure \ref{FBeffect} (solid thick).
	There is an important difference in the behavior of the two areas.
	V1 receives a weakly saturating LGN input, thus the saturation in the excitatory neurons in V1 mainly depends 
	on the recurrent inhibitory feedback, $J_{EI}^{11}$.
	In contrast, V2 receives a strongly saturated input from V1, therefore, both neuron types contrast-saturation
	is mainly determined by the feedforward input from V1.
	Excitatory CRFs in V2 are steeper than in V1 because of the transfer-function nonlinearity.

	\emph{Inter-areal feedback effects}:\\
	Assuming that attention acts by effectively modulating the inter-areal feedback synaptic strength, 
	two possible changes in the CRFs shape can occur:
	1) If changes lead to \( J_{EE}^{12} = J_{IE}^{12} \), a multiplicative contrast-gain modulation is observed (figure \ref{FBeffect}, solid and dotted).
	2) If changes lead to \( J_{EE}^{12} > J_{IE}^{12} \), a multiplicative response-gain modulation is observed (figure \ref{FBeffect}, dashed).
	
	Contrast-gain modulation effect occurs even for a small modulatory feedback, 
	whose strength is assumed to be 1/5 of the feedforward strength is shown in figure \ref{FBeffect} (solid thin).
	A further increase in the feedback from V2 to V1, keeping \(J_{EE}^{12} = J_{IE}^{12} \),
	induces a larger leftward shift in V1 excitatory CRF (figure \ref{FBeffect}, dotted).
	Contrast-gain effect in these cases is obtained because the inter-areal feedback 
	only scales the input to the neurons in V1, thus attention shifts to the right the total input to the neurons,
	while the transfer-function properties are unchanged.
	Thus, at low contrasts, the feedback provides a small additional input to V1 and therefore no changes in V1 CRFs are observed.
	At intermediate contrasts, the additional input increases the activity of both neurons for a given contrast
	relatively to the case without attention.
	However, at high contrasts inhibition becomes sufficiently strong, thus as for the single ring, 
	saturation level is mainly affected by the recurrent inhibitory input within V1.
	The additional input from V2 'pushed' the total input to the point where saturation occurs.
	Since the shifted excitatory V1 CRF serves the input to V2, 
	the CRFs of both neuron types in V2 are leftward-shifted as well.

	When \( J_{EE}^{12} > J_{IE}^{12} \) there is an additional excitatory input that drives V1 neurons 
	to higher rates for all contrast. The additional input to V1 neurons is asymmetrically 
	scales the effective input to excitatory and inhibitory neurons.
	Based on the results above, 
	the effect of attention here can be viewed as an effective change in the recurrent drive $J_{EE}^{11}$ 
	which is proportional to \( J_{EE}^{12} - J_{IE}^{12} \).
	At high contrast, similarly to the single ring, this additional excitatory input 
	recruits additional recurrent inhibition that eventually leads to saturation,
	but only for high excitatory rates. There is thus a response-gain effect.
	Note that due to symmetry of the ring and the roughly equal excitatory activity in both areas, 
	the degree of modulation in V1 and V2 are comparable.
	This indicates that similar changes in the inter-areal feedforward strength will lead 
	to the same type of modulation, as well (not shown).

	In both cases, saturation occurs over roughly the same range of contrasts.
	This is because saturation is determined mainly by the recurrent $J_{EI}$ and $J_{II}$ (Persi et al., 2008),
	thus the changes induced by excitatory drive can only lead to approximated multiplicative effects,
	in such inhibition-dominated network.
	Attentional effects for this hypothesis are significant only above 3-4\% contrast,
	because the inter-areal feedback input is proportional to the neuronal activity, $R_{E}^{2}$, 
	which is very low at low contrast.
	A clear drawback of this hypothesis is instability.
	For some effective synaptic drive, $J$, the steady state rate to a leading order 
        is \(R \sim I/(s - J) \), where $s$ is some constant smaller than 1, provided that \( \alpha > 1 \).
        Thus any mechanism that increases $J$ finally meets instability when $J$ approaches $s$ from below. \\

\begin{figure}[btp]
	  \centerline{\psfig{figure= 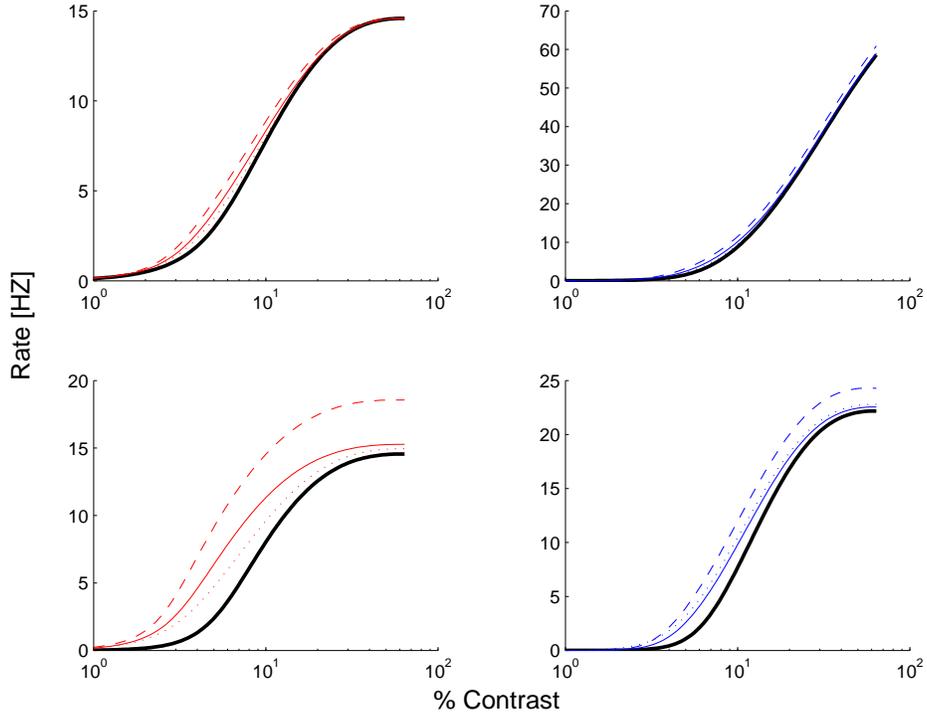,width=350pt}}
	  \caption[]{ The effect of changes in single neuron properties in V2. 
	    CRFs of excitatory (right) and inhibitory neurons (left) in V1 (upper panel) and V2 (lower panel) are shown.
	    Solid thick curves are the responses with \( J_{EE}^{21} = J_{IE}^{21} = 0.15 \), and \( J_{EE}^{12} = J_{IE}^{12} = 0.03 \)
	    and transfer-functions of \( \beta_{E} = 6.5, \ \alpha_{E} = 1.45\) \( \beta_{I} = 5.2, \ \alpha_{I} = 2.2\).
	    Solid thin curves show an approximated contrast-gain effect in the CRFs when \(\alpha_{E} \to \alpha_{E}/1.3\), and \(\beta_{E} \to 1.3\beta_{E}\).
	    Dashed curves show a response-gain effect in the CRFs when \(\alpha_{E} \to \alpha_{E}/1.3\), but \(\beta_{E} \to 1.5\beta_{E}\).
	    Dotted curves show an approximated contrast-gain effect in the CRFs when \(\alpha_{E} \to \alpha_{E}/1.3\), \(\beta_{E} \to 1.3\beta_{E}\)
	    and also \(\alpha_{I} \to \alpha_{I}/1.3\), \(\beta_{I} \to 1.5\beta_{I}\). 
	  }
	  \label{SNeffect}
\end{figure}

	\emph{Effects of changes in the effective transfer-function}:\\
	The second hypothesis assumes that the attentional signal leads to changes in the effective transfer-function.
	In principle, attention can lead to various changes in single neuron properties.
	Nevertheless, we assume here that possible changes in the transfer-function 
	by attention are induced by increasing the input noise level (see Discussion).
	An increase of noise level will tend to linearize the transfer-function (Hansel \& van Vreeswijk, 2002),
	because it affects neuronal responses more near threshold levels than at high inputs.
	We can effectively describe this effect by a decrease of the power-law exponent $\alpha$, 
	and an increase the gain $\beta$.
	Since the attentional signal is stronger in higher areas than in lower areas, 
	we will explore here the extreme case of changes only of V2 neurons.
	Therefore, we examine the effect of \(\alpha_{a} \to \alpha_{a}/A_{a} \), 
	and \(\beta_{a} \to \beta_{a}B_{a} \) in V2, on the CRFs of both V1 and V2.
	Because noise can not decrease the firing rate but only increase it,
	the constraints on $A_{a}$ and $B_{a}$ are that they must not lead 
	to a decrease in single neuron firing rate over the physiological range of firing rates 
	(20Hz for excitatory neuron, 40Hz for inhibitory neurons).
	We chose values that meet these constraints within these range of rates.
	In figure \ref{SNeffect} the responses of neurons with both feedforward and feedback connections (solid curves)
	are shown for the reference choice of $\alpha$ and $\beta$ (figure \ref{RIC}, solid).
	The effects of different $A_{a}$ and $B_{a}$ values on the CRFs are shown in figure \ref{SNeffect}.

	Similarly to the first hypothesis, both a contrast-gain and a response-gain effects can be observed.
	If changes for the excitatory neurons follow \(A_{E} = B_{E} \), a significant contrast-gain effect is observed 
	in V2 and V1, but the effect is larger in V2 (\ref{SNeffect}, solid thin).
	This is because V2 excitatory neurons are more responsive over the intermediate range of rates, 
	but not at high and low contrast and rates (see also figure \ref{RIC}, middle).
	The lower effect in V1 is a consequence of the weak inter-areal feedback connection.
	Unlike the previous case, here, attentional effects first occur in V2 and then percolate to V1.
	When \(A_{E} < B_{E} \) the excitatory neurons in V2 are more responsive over the whole range of contrasts, 
	and therefore, a response-gain effect is observed for V2, but not for V1 (\ref{SNeffect}, dashed).
	In V1, saturation is dominated by high recurrent inhibition within V1, 
	thus the weak additional input from V2 does not affect much V1's saturation level.
	Also when inhibitory neurons transfer-function properties are altered in V2,
	a contrast-gain effect is observed.
	However, for the inhibitory neurons to be more responsive over the range of rates we consider (40Hz), 
	it is necessary to chose \( A_{I} < B_{I} \) (see figure \ref{RIC}, left).
	Differently from the feedback hypothesis, linearizing the transfer-function leads to a considerable 
	attentional effect in all cases, already at very low contrasts.
	A clear drawback of this hypothesis is that changes in $\alpha$ would lead to changes in the orientation tuning width.
	However, changes in the tuning width are proportional to $\sqrt{A}$, thus for small $A$ 
	we can obtain both a considerable change in the CRF and a weak effect of a change of few degrees
	in the orientation tuning width, that may be unobserved in in-Vivo experiments
	(compare the changes in figure \ref{RIC}, and figure \ref{SNeffect}).

   \section*{Discussion}

   For early visual areas, individual CRFs can be multiplicatively modulated by attention in either a contrast-gain fashion or
   a response-gain fashion, and it is not clear which effect dominates in the population level.
   In contrast, the orientation tuning curves width remains contrast-invariant and is not affected by attention.
   Theoretical models that have been suggested as an explanation for the effects of spatial attention
   on the contrast and orientation-tuning properties of neurons in early visual areas (Hahnloser et al. 1999, 2002),
   are inconsistent with attention-invariant orientation-tuning width.

   We proposed another model which inherently captures contrast-invariant orientation tuning width,
   under certain conditions that do not impose rigid constraints (Persi et al., 2008).
   Given the lack of direct evidence for how the attentional signal affects 
   the inputs to the neurons or the properties of the neurons themselves, 
   we explored two potential mechanisms, that attention can either change the effective synaptic drive or change 
   the effective transfer-function properties of neurons, in the higher visual area.
   Nevertheless, there are several indications that these assumptions are reasonable.
   Both the observed changes in the the STAs power spectrum (Fries et al. 2001)  and in baseline activity 
   (Luck et al., 1997; Reynolds et al., 1999)
   can potentially lead to effective changes either in the transfer-function, for example by changing noise levels, 
   or in the drive to the neurons as suggested in Fries et al. (2001).
   Note that in our model changes of the effective inter-areal synaptic drive are 
   equivalent to changes in the mean input to the neurons, because all sources of inputs are tuned to the same degree in the
   orientation domain and are contrast-invariant.
   Also experiments that examined the effect of sensory interaction on attentional effects 
   (Kastner et al, 1998; Reynolds et al 1999; Reynolds \& Chelazzi 2004; Treue \& Maunsell, 1996)
   suggest that feedback form higher areas may be involved in spatial attention.

   Our results show that first, it is possible to obtain both contrast-gain and response-gain effects in both hypotheses examined.
   For the first hypothesis that attention alters inter-areal synaptic strength,
   contrast-gain effect seems to be the exception and not the rule, as it requires a certain relationship between synaptic parameters.
   In contrast, for the second hypothesis that attention effectively alters single neurons' transfer-function properties,
   contrast-gain seems to be the rule, if changes in the transfer-function properties originate in changes of the input noise level.
   Then, attention can only weakly affect the transfer-function at high contrasts, thus saturation levels should not change.
   Second, the hypothesis that attention acts by changing inter-areal feedback synaptic drive 
   has two major drawbacks: instability and lack of effect at low contrasts.
   In contrast, linearizing the input-output transfer-function leads to changes at low contrasts,
   and does not lead to instability, 
   except for the unlikely possibility that it affects inhibitory neurons much more than excitatory neurons.
   Only a small change in the power-law exponent, $\alpha$, is needed to induce significant changes in the CRFs.
   However, changes in $\alpha$ by some factor $A$ would lead to a change in the orientation tuning width
   that would be proportional to $\sqrt{A}$. 
   Thus $A$ can not be too large for this hypothesis to be consistent with the observed attention-invariant orientation-tuning width.
 
   We propose to test the possibility that attention changes the effective transfer-function properties 
   in the attended mode versus the non-attended mode.
   It can be easily checked from available data by fitting the rate-voltage to the power-law function \(R(V) = \beta(V-V_{th})_{+}^{\alpha} \),
   where $V$ in this context represents the input to the neurons.
   If attention affects the transfer-function properties, such that it effectively linearizes the transfer functions,
   we would expect that in the attended mode the power-law exponent should be smaller and the gain should be higher than
   in the non-attended mode.
   To get a further insight to the sources for this potential linearization, 
   we also propose to check if the voltage fluctuations change between the two attentional modes.
   This is because according to the model described here, linearization is a result of a change in the input noise level,
   thus larger fluctuations are predicted in the attended mode.

   \subsection*{Acknowledgment}

   I would like to thank my supervisor Carl van Vreeswijk for guiding me throughout this work.
 
  \newpage
  
  \small

     \section*{References}

     Albrecht D.G., Hamilton D.B., 1982.
     Striate cortex of monkey and Cat: Contrast response function.
     {\em Journal of neurophysiology vol. 48,} { 217-237}.\\

     Anderson J.S., Lampl I., Gillespie D.C. Frester D., 2000.
     The contribution of noise to contrast invariance of orientation tuning in cat visual cortex.
     {\em Science vol. 290,} { 1968-1972}.\\
     
     Ben-Yishai, Lev Bar-Or and Sompolinsky H., 1995.
     Theory of orientation tuning in visual cortex.
     {\em Neurobiology vol. 92,} { 3844-3848}.\\

     Boynton G.M.,  Buracas G.T., 2007.
     The effect of spatial attention on contrast response functions in human visual cortex.
     {\em Journal of Neuroscience vol. 27(1):} { 93-97}.\\

     Contreras D., and Palmer L., 2003.
     Response to contrast of electro-physiologically defined cell classes in primary visual cortex.
     {\em Journal of neuroscience vol. 23(17),} { 6936-6945}.\\

     Finn I.M, Priebe N.J, and Ferster D., 2007.
     The emergence of contrast-invariant orientation tuning in simple cells of cat visual cortex.
     {\em Neuron vol. 54,} {137-152}.\\

     Fries P, Reynolds JH, Rorie AE, Desimone R, 2001.
     Modulation of oscillatory neuronal synchronization by selective visual attention.
     {\em Science vol. 291,} {1560-1562}.\\

     Hahnloser R., Douglas R.J., Mahowald M., Hepp K., 1999.
     Feedback interactions between neuronal pointers and maps for attentional processing.
     {\em Nature Neuroscience vol 2(8):} {746-752}.\\

     Hahnloser R., Douglas R.J., Hepp K., 2002.
     Attentional recruitment of inter-areal recurrent networks for selective gain control.
     {\em Neural Computation vol 14:} {1669-1689}.\\
     
     Hansel D. and van Vreeswijk C., 2002.
     How noise contributes to contrast invariance of orientation tuning in cat visual cortex.
     {\em Journal of neuroscience vol. 22(12),} { 5118-5128}.\\

     Hopfinger J.B., Buonocore M.H., Mangum G.R., 2000.
     The neuronal mechanisms of top-down attentional control.
     {\em Nature Neuroscience vol. 3(3),} { 284-291}.\\

     Kastner S., Weerd D.P., Desimone R., Ungerleider G.L., 1998.
     Mechanisms of directed attention in the human extrastriate cortex as revealed by functional MRI.
     {\em Science, vol 282:} { 108-111}.\\

     Luck J.S., Chelazzi L., Hillyard A.S., Desimone R., 1997.
     Neural mechanisms of spatial selective attention in areas V1, V2, V4 of macaque visual cortex.
     {\em Journal of Neurophysiology vol. 77,} { 24-42}.\\
      
     McAdams J.C., Maunsell J.H.R.., 1999.
     Effects of attention on orientation tuning curve functions of single neurons in macaque cortical area V4.
     {\em Journal of Neuroscience vol. 19(1),} { 431-441}.\\ 

     Miller K.D., Troyer W.T., 2002.
     Neural noise can explain expansive, power-law nonlinearities in neural response functions.
     {\em Journal of Neurophysiology. vol 87(2),} { 653-9}.\\

     Pessoa L., Kastner S., Ungerleider L.G., 2003.
     Neuroimaging studies of attention: From modulation of sensory processing to top-down control.
     {\em Journal of Neuroscience 23(10):} { 3990-3998}.\\

     Persi E., Hansel D., Nowak L., Barone P., van Vreeswijk C., 2008.
     Power-Law input-output transfer functions explain the contrast response and tuning properties of neurons in the visual cortex.
     {\em Frontiers in computational Neuroscience} { submitted for publication}.\\

     Priebe N.J., Mechler F., Carandini M., Ferster D., 2004.
     The contribution of spike threshold to the dichotomy pf cortical simple and complex cells.
     {\em Nature Neuroscience vol. 7(10),} {1113-1122}. \\

     Reynolds J.H., Chelazzi L, 2004.
     Attentional modulation of visual processing.
     {\em Annu. Rev. Neuroscience vol 27:} { 611-647}.\\

     Reynolds J.H., Chelazzi L, Desimone R, 1999.
     Competitive mechanisms subserve attention in Macaque areas V2 and V4.
     {\em Journal of Neuroscience vol 19(5)} { 1736-1753}.\\
 
     Reynolds J.H., Pasternak P, Desimone R., 2000.
     Attention increases sensitivity of V4 neurons.
     {\em Neuron vol. 26,} { 703-714}.\\

     Sclar G., Freeman R., 1982.
     Orientation selectivity in cat's striate cortex is invariant with stimulus contrast.
     {\em Exp. Brain Res. vol. 46,} { 457-461}.\\
     
     Sclar G, Maunsell J.H., and Lennie P., 1990.
     Coding of image contrast in central visual pathways of the macaque monkey.
     {\em Vision. vol. 30,} { 1-10}.\\    

     Skottun B.C., Bradley A., Sclar G., Ohzawa I., Freeman R.D., 1987.
     The effects of contrast on visual orientation and spatial frequency discrimination:
     A comparison of single cells behavior.
     {\em Journal of Neurophysiology, 57(3),} { 773-786}.\\

     Treue S, Maunsell J.H.R., 1996.
     Attentional modulation of visual motion processing in cortical areas MT and MST.
     {\em Nature, vol. 382:} { 539-541}.\\

     Williford T., Maunsell J.H.R., 2006.
     Effects of spatial attention on contrast response functions in Macaque area V4.
     {\em Journal of Neurophysiology vol. 96:} { 40-54}.\\

\newpage

\end{document}